\documentclass[epj]{webofc}
\usepackage[varg]{txfonts}   
\usepackage{amssymb}
\usepackage{amsmath}
\usepackage{hyperref}

\def\P{{\boldsymbol P}}

\newcommand{\der}{\mathrm{d}}
\newcommand{\xt}{{{\boldsymbol x}_\perp}}
\newcommand{\yt}{{{\boldsymbol y}_\perp}}
\newcommand{\bt}{{{\boldsymbol b}_\perp}}
\newcommand{\rt}{{{\boldsymbol r}_\perp}}

\newcommand{\Pt}{{\P_\perp}}

\newcommand{\ud}{\, \mathrm{d}}
\newcommand{\tr}{\, \mathrm{Tr} \, }
\newcommand{\nc}{{N_\mathrm{c}}}

\newcommand{\qso}{Q_\mathrm{s0}}

\newcommand{\lqcd}{\Lambda_{\mathrm{QCD}}}
\newcommand{\as}{\alpha_{\mathrm{s}}}

\newcommand{\Jpsi}{{J/\psi}}

\woctitle{Physics Opportunities at an Electron-Ion Collider}
\begin{document}
\title{Centrality-dependent forward $J/\psi$ production in high energy proton-nucleus collisions}
%
%

\author{B. Duclou\'e\inst{1,2} 
\and T. Lappi\inst{1,2} 
\and H. M\"antysaari\inst{3} 
}

\institute{Department of Physics, University of Jyv\"askyl\"a, P.O. Box 35, 40014 University of Jyv\"askyl\"a, Finland
\and Helsinki Institute of Physics, P.O. Box 64, 00014 University of Helsinki, Finland
\and Physics Department, Brookhaven National Laboratory, Upton, NY 11973, USA
}

\abstract{Forward $J/\psi$ production and suppression in high energy proton-nucleus collisions can be an important probe of gluon saturation. In an earlier work we studied this process in the Color Glass Condensate framework and showed that using the Glauber approach to extrapolate the dipole cross section of a proton to a nucleus leads to results closer to experimental data than previous calculations in this framework. Here we investigate the centrality dependence of the nuclear suppression in this model and show a comparison of our results with recent LHC data.}
\maketitle
\section{Introduction}

The study of the nuclear suppression of forward $J/\psi$ production in high energy proton-nucleus collisions can be a valuable tool to better understand saturation dynamics. Indeed, it probes the target nucleus at very high densities, where saturation effects should be enhanced, and the charm quark mass is
small enough to be sensitive to these dynamics while being large enough to provide a hard scale allowing a perturbative treatment. $J/\psi$ mesons also have clean experimental signatures and their production and suppression have been the subject of many experimental studies.
In a recent work~\cite{Ducloue:2015gfa} we re-evaluated the nuclear suppression of forward $J/\psi$ production at high energy in the Color Glass Condensate (CGC) framework, showing that using the optical Glauber model to relate the dipole cross section of a nucleus to the one of a proton leads to a smaller suppression for minimum bias events than in previous works in the same formalism~\cite{Fujii:2013gxa} and results closer to experimental data\footnote{The authors of~\cite{Fujii:2013gxa} have recently presented updated results~\cite{Fujii:2015lld} similar to those obtained in~\cite{Ducloue:2015gfa}.}.
In this work we discuss the relation between the explicit impact parameter dependence of the optical Glauber model and centrality and we compare our results with recent data on the centrality dependence of $J/\psi$ suppression at the LHC presented by the ALICE Collaboration~\cite{Adam:2015jsa}.

\section{Formalism}

The formalism for gluon and quark pair production in the dilute-dense limit of the Color Glass Condensate has been studied in Refs.~\cite{Blaizot:2004wu,Blaizot:2004wv} (see also Ref.~\cite{Kharzeev:2012py}) and applied in several works, such as~\cite{Fujii:2005rm,Fujii:2006ab,Fujii:2013gxa,Fujii:2013yja,Ma:2015sia}.
This allows to compute the cross section for $c\bar{c}$ pair production, which is the central object needed to study $J/\psi$ production. The expression for this cross section can be found in Ref.~\cite{Fujii:2013gxa}.
We use the simple color evaporation model to describe the hadronization of the produced $c\bar{c}$ pairs in $J/\psi$ mesons. In this model, a fixed fraction of the $c\bar{c}$ pairs produced with an invariant mass between $2m_c$ and $2M_D$, where $m_c$ is the charm quark mass and $m_D$ is the $D$-meson mass, is assumed to hadronize into $\Jpsi$ mesons.
Thus the cross section for $J/\psi$ production with transverse momentum $\P_{\perp}$ and rapidity $Y$ reads
\begin{align} 
	\frac{\ud\sigma_{\Jpsi}}{\ud^2\P_{\perp}\ud Y}
	=
	F_{\Jpsi} \; \int_{4m_c^2}^{4M_D^2} \ud M^2
	\frac{\ud\sigma_{c\bar c}}
	{\ud^2\P_{\perp} \ud Y \ud M^2}
	\, ,
	\label{eq:dsigmajpsi}
\end{align}
where $\frac{\ud\sigma_{c\bar c}}{\ud^2\P_{\perp} \ud Y \ud M^2}$ is the cross section for the production of a $c\bar{c}$ pair with transverse momentum $\P_{\perp}$, rapidity $Y$ and invariant mass $M$. In this expression $F_{\Jpsi}$ is a non-perturbative constant which can be extracted from data. In the following we will focus on ratios of cross sections for which the value of this parameter plays no role.

In forward $J/\psi$ production, the projectile proton is probed at relatively large $x$ which justifies the use of the collinear approximation on this side. The gluon density in the projectile can thus be described using a collinear parton distribution function. Here we will use the MSTW 2008~\cite{Martin:2009iq} LO parametrization for this purpose.
On the other hand, when working at forward rapidity, the target, which can be either a proton or a nucleus, is probed at very small $x$. The information about its gluon density is contained in the function
\begin{equation}
	S_{_Y}(\xt-\yt) = \frac{1}{\nc }\left< \tr U^\dag(\xt)U(\yt)\right>,
\end{equation}
where $U(\xt)$ is a fundamental representation Wilson line in the target color field.
The evolution of $S_{_Y}(\rt)$ is governed by the running coupling Balitsky-Kovchegov (rcBK) equation~\cite{Balitsky:1995ub,Kovchegov:1999ua,Balitsky:2006wa}, which is solved numerically. We use as an initial condition the MV$^e$ parametrization~\cite{Lappi:2013zma} which involves, as the AAMQS~\cite{Albacete:2010sy} one, parameters which are extracted from DIS measurements.
In the MV$^e$ parametrization the initial condition for a proton target reads 
\begin{equation}\label{eq:icp}
S^p_{Y= \ln \frac{1}{x_0}}(\rt) = \exp \bigg[ -\frac{\rt^2 \qso^2}{4} 
\ln \left(\frac{1}{|\rt| \lqcd}\!+\!e_c \cdot e\right)\bigg] \, ,
\end{equation}
with $x_0=0.01$. The expression for the running coupling is
\begin{equation}
\as(r) = \frac{12\pi}{(33 - 2N_f) \log \left(\frac{4C^2}{r^2\lqcd^2} \right)} \, .
\end{equation}
In this case there is no explicit dependence on the impact parameter and the integration over $\bt$ can be simply replaced by
\begin{equation}\label{eq:defsigma0}
\int\der^2 \bt \to \frac{\sigma_0}{2} \, ,
\end{equation}
where $\frac{\sigma_0}{2}$ corresponds to the effective proton transverse area measured in DIS experiments. The values of the parameters in these expressions obtained in Ref.~\cite{Lappi:2013zma} by a fit to HERA DIS data~\cite{Aaron:2009aa} are $\qso^2= 0.060$ GeV$^2$, $C^2= 7.2$, $e_c=18.9$  and $\frac{\sigma_0}{2} = 16.36$ mb.
To generalize the dipole cross section from a proton to a nucleus we use, as in Ref.~\cite{Lappi:2013zma}, the optical Glauber model. In this approach the initial condition to the rcBK equation reads
\begin{equation}\label{eq:ica}
S^A_{Y=\ln \frac{1}{x_0}}(\rt,\bt) = \exp\Bigg[ -A T_A(\bt) 
\frac{\sigma_0}{2} \frac{\rt^2 \qso^2}{4} 
\ln \left(\frac{1}{|\rt|\lqcd}+e_c \cdot e\right) \Bigg] \; ,
\end{equation}
where the only additional quantity involved compared to the proton case~(\ref{eq:icp}) is the standard Woods-Saxon distribution $T_A(\bt)$, where $\bt$ is the impact parameter:
\begin{equation}
T_A(\bt)= \int \der z \frac{n}{1+\exp \left[ \frac{\sqrt{\bt^2 + z^2}-R_A}{d} \right]} \; ,
\end{equation}
with $d=0.54\,\mathrm{fm}$, $R_A=(1.12A^{1/3}-0.86A^{-1/3})\,\mathrm{fm}$ and $n$ is a normalization constant defined such that $\int \der^2\bt T_A(\bt)=1$. Besides $T_A$, all the parameters take the same value as in the proton case. The rcBK equation is then solved independently for each value of $\bt$. In principle this explicit impact parameter dependence can be related to the centrality classes used by experiments in the following way: in the optical Glauber model a class $(c_1-c_2)\%$ would be defined by limiting the integration over $\bt$ between $b_1$ and $b_2$ defined such that
\begin{equation}
(c_1-c_2)\% = \frac{1}{\sigma_\text{inel}^{\text{pA}}} \int_{b_1}^{b_2} \der^2 \bt p(\bt).
\label{eq:centrality}
\end{equation}
Here $\sigma_\text{inel}^{\text{pA}}$ is the total inelastic proton-nucleus cross section, given by
\begin{equation}
\sigma_\text{inel}^{\text{pA}} = \int \der^2 \bt \, p(\bt) \; ,
\end{equation}	
with the scattering probability at impact parameter $\bt$ being
\begin{equation}
p(\bt) \approx 1- e^{-A T_A(\bt) \sigma_\text{inel}},
\end{equation}
where $\sigma_\text{inel}$ is the total inelastic nucleon-nucleon cross section. The particle yield in a given centrality class is then given by
\begin{equation}
\frac{\der N}{\der^2 \Pt \der Y} = \frac{ \int_{b_1}^{b_2} \der^2 \bt \frac{\der N(\bt)}{\der^2 \Pt \der Y} } {\int_{b_1}^{b_2} \der^2 \bt \, p(\bt)},
\end{equation}
where the values of $b_1$ and $b_2$ are defined as in Eq.~(\ref{eq:centrality}).
However, if we use this procedure and compare the values of the average number of binary nucleon-nucleon collisions, given in this model by
\begin{equation}
\langle N_\text{coll} \rangle_\text{opt.} = \frac{\int_{b_1}^{b_2} \der^2 \bt N_\text{coll opt.}(\bt)}{\int_{b_1}^{b_2} \der^2 \bt p(\bt)},
\end{equation}
where
\begin{equation}
N_\text{coll opt.}(\bt) = A T_A(\bt) \sigma_\text{inel} \,,
\end{equation}
with the values of $\langle N_\text{coll} \rangle$ estimated by ALICE~\cite{Adam:2015jsa} in each centrality class, we find a disagreement between the two as can be seen from Table~\ref{tab:Ncoll}. In the following, for a first comparison, we decide to perform our calculation in each centrality class at a fixed impact parameter $\bt$ defined such that the number of binary collisions in the optical Glauber model corresponds to its average value estimated by ALICE, i.e. $N_\text{coll opt.}(\bt)=\langle N_\text{coll} \rangle_\text{ALICE}$. A more consistent comparison would require the use of distributions in impact parameter space but for this one would need to have access to the $N_\text{coll}$ distributions at experiments and not only to $\langle N_\text{coll} \rangle$.

\begin{table}[h]
	\centering\begin{tabular}{|c|c|c|}
		\hline
		Centrality class & $\langle N_\text{coll} \rangle_\text{opt.}$ & $\langle N_\text{coll} \rangle_\text{ALICE}$ \\
		\hline
		2--10\%   & 14.7 & $11.7 \pm 1.2 \pm 0.9$\\
        10--20\%  & 13.6 & $11.0 \pm 0.4 \pm 0.9$\\
		20--40\%  & 11.4 & $9.6 \pm 0.2 \pm 0.8$ \\
		40--60\%  & 7.7  & $7.1 \pm 0.3 \pm 0.6$ \\
		60--80\%  & 3.7  & $4.3 \pm 0.3 \pm 0.3$ \\
		80--100\% & 1.5  & $2.1 \pm 0.1 \pm 0.2$ \\
		\hline
	\end{tabular}
	\caption{Average number of binary collisions in each centrality class as obtained in the optical Glauber model compared with the value estimated by ALICE~\cite{Adam:2015jsa}.}
	\label{tab:Ncoll}
\end{table}

\begin{figure}[h]
	\def\picheight{4cm}
	\centering\includegraphics[height=\picheight]{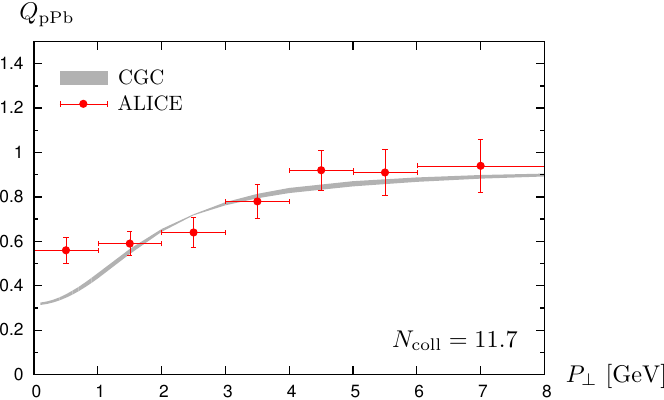}
	\hspace{0.45cm}
	\includegraphics[height=\picheight]{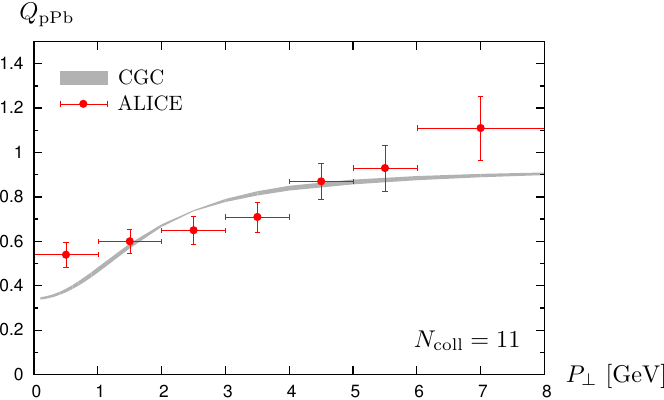}
	
	\vspace{0.2cm}
	\includegraphics[height=\picheight]{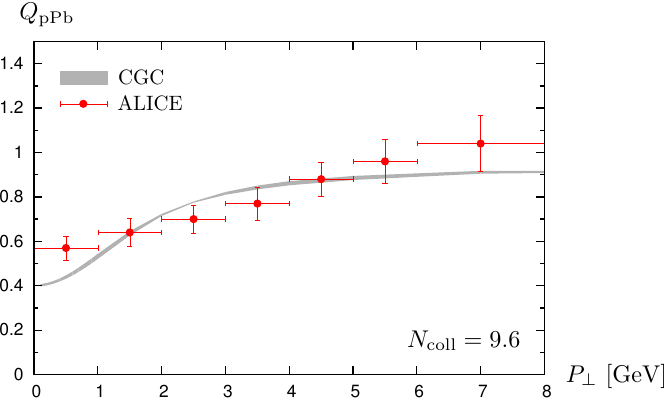}
	\hspace{0.4cm}
	\includegraphics[height=\picheight]{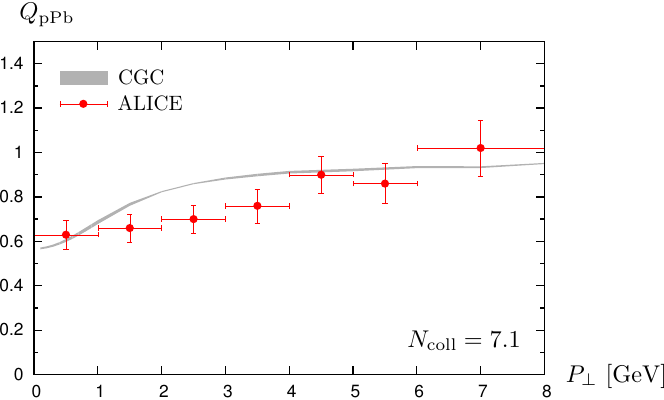}
	
	\vspace{0.2cm}
	\includegraphics[height=\picheight]{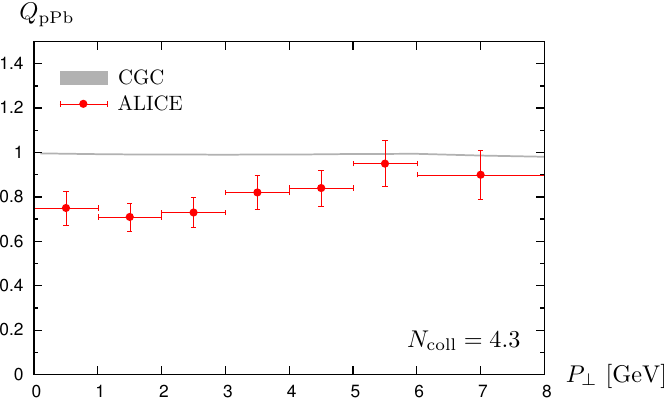}
	\caption{Nuclear modification factor $Q_\text{pPb}$ as a function of $P_\perp$ in different centrality bins compared with ALICE data~\cite{Adam:2015jsa}.}
	\label{fig:QpA_bins}
\end{figure}

\begin{figure}[h]
	\def\picheight{4.5cm}
	\centering\includegraphics[height=\picheight]{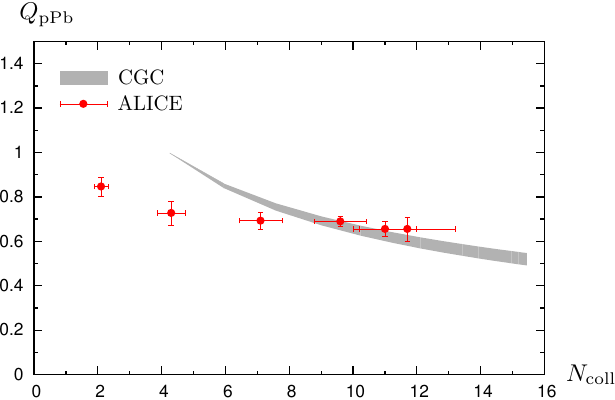}
	\hspace{0.4cm}
	\includegraphics[height=\picheight]{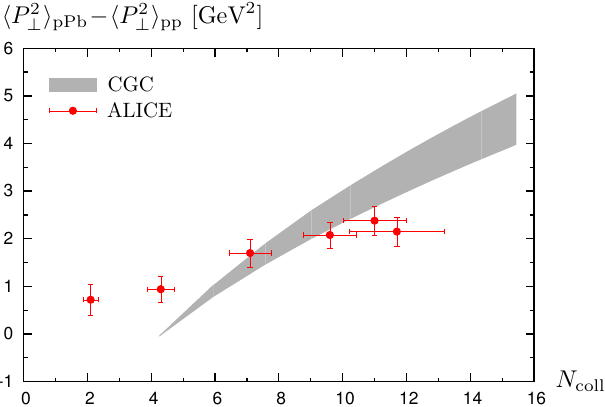}
	\caption{Nuclear modification factor $Q_\text{pPb}$ (left) and nuclear transverse momentum broadening $\langle P_\perp^2 \rangle_\text{pPb}-\langle P_\perp^2 \rangle_\text{pp}$ (right) as a function of $N_\text{coll}$ compared with ALICE data~\cite{Adam:2015jsa}.}
	\label{fig:QpA_deltapT2_Ncoll}
\end{figure}

\section{Results}

The ALICE collaboration recently measured the nuclear suppression of $J/\psi$ production in proton-lead collisions in different centrality classes at $\sqrt{s_{NN}}=5$ TeV~\cite{Adam:2015jsa}. To compare our results with these data we use, as explained previously, a fixed impact parameter defined such that $N_\text{coll. opt}(\bt)=\langle N_\text{coll} \rangle_\text{ALICE}$ in each centrality class considered by ALICE. This procedure would lead for the 80--100\% class to an impact parameter for which the saturation scale of the nucleus would fall below the one of the proton. For this reason we will not consider this class in the following.
In Fig.~\ref{fig:QpA_bins} we show the comparison of our results and ALICE data for the nuclear modification factor $Q_\text{pPb}$, defined as
\begin{equation}
Q_{\rm pPb}= \frac{\frac{\ud N^\text{pPb}}{\ud^2 \Pt \ud Y}}
{ A \langle T_A \rangle \frac{\ud\sigma^\text{pp}}{\ud^2 \Pt \ud Y}} \; ,
\end{equation}
as a function of $P_\perp$ in the five most central classes considered by ALICE. We include in the uncertainty band the variation of $m_c$ between 1.2 and 1.5 GeV and of the factorization scale between $M_\perp/2$ and $2M_\perp$ with $M_\perp=\sqrt{M^2+P_{\perp}^2}$ where $M$ is the $c\bar{c}$ pair's invariant mass. The description of the data is quite good in the first three bins but our calculation predicts values of $Q_\text{pPb}$ which approach unity too quickly when $N_\text{coll}$ decreases.
This too strong dependence on centrality in our calculation can also be seen in Fig.~\ref{fig:QpA_deltapT2_Ncoll}, where we show the nuclear modification factor integrated over $P_\perp$ as well as the transverse momentum broadening, defined as the difference of $\langle P_\perp^2 \rangle$ in proton-lead and in proton-proton collisions, as a function $N_\text{coll}$. However, one should keep in mind that the value of $N_\text{coll}$ indicated for ALICE data is an average while in our calculation it is a fixed value. Taking into account the fluctuations in our calculation could have a significant impact.

\section{Conclusions}

In this work we have studied the centrality dependence of forward $J/\psi$ nuclear suppression in the Color Glass Condensate. For this we used, as in Ref.~\cite{Ducloue:2015gfa}, the optical Glauber model to generalize the dipole cross section of a proton to nuclei. In Ref.~\cite{Ducloue:2015gfa} this model was found to lead to values for the nuclear modification factor in minimum bias collisions closer to experimental results than previous calculations in a similar framework. However, when studying the centrality dependence in this model, we found here that this dependence appears to be much stronger than in recent ALICE data~\cite{Adam:2015jsa}. Nevertheless, we stress that the results shown here have been obtained using a fixed impact parameter for which, in the optical Glauber model, the number of binary collisions corresponds to the average value of this quantity estimated by ALICE. For a more consistent comparison with experimental data, it would be necessary to use distributions in impact parameter space, but this would require access to the typical size of the fluctuations of the number of binary collisions in the experimental centrality classes.

\section*{Acknowledgments}
T.~L. and B.~D. are supported by the Academy of Finland, projects
267321 and 273464. 
H.~M. is supported under DOE Contract No. DE-SC0012704.
This research used computing resources of 
CSC -- IT Center for Science in Espoo, Finland.
We would like to thank C. Hadjidakis and I. Lakomov for 
discussions on the ALICE data.

\end{document}